# Pressure-driven switching of magnetism in layered CrCl$_3$

Azkar Saeed Ahmad[1], Yongcheng Liang[2], Mingdong Dong[1], Xuefeng Zhou[1], Leiming Fang[3], Yuanhua Xia[3], Jianhong Dai[1,4], Xiaozhi Yan[1,4], Xiaohui Yu[4], Guojun Zhang[2], Yusheng Zhao[1] & Shanmin Wang[1]

[1]Department of Physics and Shenzhen Engineering Research Center of Frontier Materials Synthesis at High Pressure, Southern University of Science and Technology, Guangdong 518055, China. [2]College of Science, Institute of Functional Materials, and State Key Laboratory for Modification of Chemical Fibers and Polymer Materials, Donghua University, Shanghai 201620, China. [3]Key Laboratory for Neutron Physics, Institute of Nuclear Physics and Chemistry, China Academy of Engineering Physics, Mianyang 621999, China. [4]Key Laboratory of Extreme Conditions Physics, Institute of Physics, Chinese Academy of Sciences, Beijing 100190,China.

Correspondence and requests for materials should be addressed to S.W. (email: wangsm@sustech.edu.cn) or to Y.L. (email: ycliang@dhu.edu.cn)

Layered transition-metal compounds with controllable magnetic behaviors provide many fascinating opportunities for the fabrication of high-performance magneto-electric and spintronic devices. The tuning of their electronic and magnetic properties is usually limited to the change of layer thickness, electrostatic doping, and the control of electric and magnetic fields. However, pressure has been rarely exploited as a control parameter for tailoring their magneto-electric properties. Here, we report a unique pressure-driven isostructural phase transition in layered CrCl$_3$ accompanied by a simultaneous switching of magnetism from a ferromagnetic to an antiferromagnetic ordering. Our experiments, in combination with *ab initio* calculations, demonstrate that such a magnetic transformation hinders the band-gap collapse under pressure, leading to an anomalous semiconductor-to-semiconductor transition. Our findings not only reveal high potential of CrCl$_3$ in electronic and spintronic applications under ambient and extreme conditions but also establish the basis for exploring unusual phase transitions in layered transition-metal compounds.

Layered transition-metal (TM) compounds are a large class of important functional materials and have been a focus of recent research because of their intriguing optical, electronic, and magnetic properties[1-7]. As typical examples, single- or multi-layered TM dichalcogenides are identified as promising materials for valleytronics, spintronics, and optoelectronics devices[8-10]. These functionalities are benefited from their unique layered structural configurations. The most prominent feature in the crystal structure is that one sheet of hexagonally packed TM atoms are sandwiched by two sheets of anions. These sandwich layers are vertically stacked and loosely linked by weak van der Waals (vdW) force, which structurally facilitates the exfoliation of their three-dimensional (3D) bulk counterpart into two-dimensional (2D) layers. This has led to the discovery of many appealing properties in the atomic thickness of 2D materials[11-15]. Unlike graphene, lots of 2D TM compounds possess desirable electronic and magnetic properties, making them promising candidates for the fabrication of low-power and ultra-thin transistors to achieve higher work efficiency than conventional silicon-based transistors[16-18].

Among these layered TM compounds, chromium trihalides ($CrX_3$, X=Cl, Br, and I) are of particular interest due to their extraordinary electronic and magnetic properties[19-25]. In each layer, Cr and X atoms are arranged in a hexagonal network, in which one Cr atom is bonded to its six neighboring X atoms to form edge-shared octahedra. The resulting honeycomb patterns of X-Cr-X are stacked along the *c*-axis and are held by weak vdW interactions. This local coordination environment makes the Cr-$t_{2g}$ orbitals half-filled, which creates complex magnetostructural behaviors in $CrX_3$, similar to ternary layered $CrGeTe_3$ and $CrSiTe_3$[26]. It is reported that $CrX_3$ are paramagnetic (PM) at room temperature, but their magnetisms are different at low temperature. Both $CrI_3$ and $CrBr_3$ are ferromagnetic (FM) with the Curie temperature of 61 K[23] and 37 K[27], respectively, whereas $CrCl_3$ is antiferromagnetic (AFM) with the Néel temperature of 16.8 K[28]. On the other hand, the structural transitions from the rhombohedral to monoclinic symmetry are also observed at 210 K for $CrI_3$[23], 230 K for $CrCl_3$[19,24], and 420 K for $CrBr_3$[19,20,23]. Although these reported crystallographic transitions do not involve any magnetic changes, some observed anomalies in the magnetic susceptibility at the crystallographic transition of $CrCl_3$ and in the lattice spacing at the Curie temperature of $CrI_3$ indicate the coupling between the lattice and the magnetism[23,24]. Besides the magnetostructural phase transitions, $CrX_3$ possess favorable band gaps, low cleavable energy, and tunable magnetic properties, which offer great promise for the fabrication of the next-generation 2D magneto-optical, magneto-electric, and spintronic devices.

In order to realize the potential functionalities of $CrX_3$, several techniques, such as changing layer thickness, the electrostatic doping, and controlling external electric and magnetic fields, have been used to tune their magneto-electric performances[29-31]. Aside from these non-pressure techniques, pressure can be an effective tool for tuning their versatile electronic and magnetic properties away from the pristine states. For example, pressure has been recently shown to tune the electronic bands in TM dichalcogenides (e.g., $MoS_2$, $MoSe_2$, and $WSe_2$)[32-38], twisted bilayer graphene[39], and graphene moiré superlattices[40]. Moreover, pressure has been also observed to alter critical temperatures of vdW magnets[41,42]. However, how pressure affects the magneto-electric properties of $CrX_3$ is yet to be explored, and this has impeded in-depth understanding and further exploration of these promising materials.

In this work, we present a high-pressure (HP) study of structural, magnetic, and electronic properties of $CrCl_3$, a prototype representative of $CrX_3$, using neutron diffraction, synchrotron X-ray diffractions (XRD), Raman scattering, and photoluminescence (PL) measurements, in combination with *ab initio* calculations. We demonstrate that $CrCl_3$ undergoes a pressured-induced switching of magnetism from a FM to an AFM ordering without involving structural symmetry breaking. Equally importantly, such a unique magnetic crossover hinders the band-gap closuring of this system under pressure, giving rise to an unusual semiconductor-to-semiconductor transition.

**Results and Discussion**

The neutron diffraction experiments have been conducted to determine the crystal structures of $CrCl_3$ at different temperatures, and our collected diffraction patterns are shown in Fig. 1 (also see Supplementary Fig. 1). At 293 K, the diffraction data can be refined with a monoclinic structure (space group $C2/m$). Its unit cell contains four $CrCl_3$ formula units, in which four Cr atoms occupy the 4$g$ (0, 0.1688, 0) Wyckoff site while twelve Cl atoms sit the 4$i$ (0.2187, 0, 0.2280) and 8$j$ (0.2497, 0.3250, 0.2259) positions. Upon cooling, a number of new diffraction peaks appear at ~240 K, suggesting a structural phase transition. Based on our refinement, this low-temperature phase can be well indexed with a rhombohedral structure (space group $R$-3) without the presence of additional magnetic peaks, implying the decoupling of structural and magnetic transitions. A careful analysis of the patterns indicates that a small fraction of monoclinic phase remains down to 100 K and coexists with the rhombohedral phase. Table 1 lists our refined lattice parameters, in good agreement with the available data[19,24,28]. According to the reported single-crystal neutron diffractions, the rhombohedral phase undergoes a magnetic transition from the PM to the AFM ordering at 17 K[28]. Moreover, McGuire et al.[24] have recently identified that the AFM ground state can be altered to the FM state by controlling the external magnetic fields. Whether these desirable magnetic switching behaviors can be manipulated by certain techniques is worthy of research, and pressure is likely to be an effective technique tool for tuning magnetic properties.

We have performed *in-situ* synchrotron XRD measurements to examine the effect of pressure on the structural stability of $CrCl_3$. At 300 K, our measured XRD patterns at different pressures are displayed in

Fig. 2a. As expected, the XRD peaks shift towards higher $2\theta$ values (i.e., low $d$-spacing) with increasing pressure. Some peaks (e.g., 020, 110, 040, and 11-3) are asymmetrically broadened. At the same time, the intensities of the 130, 200, and 113 peaks decrease gradually and almost vanish above ~20 GPa, which should be attributed to the interlayer stacking faults in $CrCl_3$. The similar phenomenon has also been observed in other 2D materials such as $RuCl_3$[43], $MoS_2$[44], and $MoN_2$[45]. Up to 40 GPa, all XRD patterns of $CrCl_3$ are indexed to the monoclinic phase, and there are no signs of symmetry changes. However, an apparently small but conceivable volume discontinuity happens at ~11 GPa, indicating a pressure-induced isostructural phase transition (Fig. 2b). The obtained pressure-volume data can be well fitted with two independent second-order Birch-Murnaghan equation of states below and above 11 GPa. The derived bulk moduli of the low-pressure (LP) and HP phases are 28(2) and 36(3) GPa, respectively. Evidently, it is reasonable that the LP phase is more compressible than the HP phase. A further analysis of the variation of axis ratios (i.e., $a/c$ and $b/c$) with pressure also confirms this isostructural transition (Fig. 2b). Below 11 GPa, as pressure rises, the axis ratio $a/c$ decreases whereas $b/c$ increases. Above 11 GPa, both $a/c$ and $b/c$ are gradually leveled off. Interestingly, the isostructural transition is reversible upon decompression (Supplementary Fig. 2).

High-pressure Raman measurements have been performed to reveal the isostructural phase transition in $CrCl_3$. As seen in Fig. 3a (and also Supplementary Fig. 3a), its room-temperature Raman spectra are characterized with six phonon modes denoted as $M1$, $M2$, $M3$, $M4$, $M5$, and $M6$, respectively, corresponding to optical branches near the center of Brillion zone. The derived frequencies of each phonon mode at HP are plotted in Fig. 3b ( and also Supplementary Fig. 3b). Upon compression, all the modes move towards higher frequencies since the optical mode is substantially enhanced at pressures. This is not unexpected since the external pressure acts as additional restoring force and exerts on the out-of-phase vibrations (i.e., optical modes), which turns out to increase the frequencies of optical modes[46]. Interestingly, the relative intensity of the $M2$ to that of $M3$ mode, $I_{M2}/I_{M3}$, appreciably increases with increasing pressure, while the separation between $M2$ and $M3$ lines is reduced and merged into one mode at ~10 GPa (Fig. 3 and Supplementary Fig. 4), similar to previous reports in binary semiconductors including ZnTe[47] and BP[48]. This phenomenon can be interpreted in terms of the electro-optical effect and the variation of effective charge at HP. Accordingly, the $M3$ and $M2$ modes should correspond to the longitudinal and transverse optical modes. Also, the pressure at which $M3$ and $M2$ modes merge together corresponds to the critical pressure for the isostructural transition as observed in the XRD experiments (Fig. 2b). This suggests that the isostructural transition would be closely associated with phonon transition. Further work is warranted to clarify this interesting isostructural transition in $CrCl_3$ from the point of view of phonon degree of freedom at pressures.

To gain insight into the phonons of $CrCl_3$, the Raman peak width and phonon mode separation are also analyzed and depicted in Figs. 3c-d, respectively, both of which can also be divided into two distinct regions with a critical pressure of ~10 GPa. This is compatible with to that of phonon frequencies as seen in Fig. 3b, further confirming that the pressure-induced phonon changes strongly correlate to the structural transition. In Fig. 3d, a drastic increase of the widths of Raman lines above ~10 GPa would be related with a stress-induced phenomenon because the HP phase is mechanically stiffer than the LP phase and builds up the large stress in the samples. These observations are in excellent agreement with our XRD measurements (Fig. 2b). Upon decompression the isostructural phase transition is again found to be reversible from Raman spectroscopy (Supplementary Fig. 4b).

High-pressure PL has been conducted to explore the variation of optical properties of $CrCl_3$ at room temperature, as shown in Figs. 4a-b (and also Supplementary Figs. 5-6). Remarkably, the PL spectrum is mainly characterized with an anomalously broadened peak centered at ~844 nm at ambient conditions. On a close inspection, this broadened peak consists of three components located around 820, 850 and 880 nm, which infers an indirect band gap in $CrCl_3$ and is evidenced by our theoretical simulations (see below). At the initial stage of compression, these PL peaks progressively move to the low-wavelength side and the width of the peaks are slightly reduced with increasing pressure up to ~11 GPa. Interestingly, with the increase of pressure above ~11 GPa the PL peaks shift backwards to the high wavelength along with significant peak broadening. In fact, the PL spectrum of a substance is often intimately associated with its indirect band gap. As commonly accepted, the band gap can be derived on the basis of PL spectrum, using its high-energy (i.e., low-wavelength) edge (Supplementary Fig. 5a). Accordingly, the derived band gap for $CrCl_3$ is plotted in Fig. 4b. As pressure increases, the band gap is anomalously increased from 1.67 eV at ambient pressure to the peak value of 1.8 eV at ~11 GPa, which indicates that the localization of the Cr-3$d$ electrons is largely enhanced in the LP phase at pressures. Above 11 GPa, the band gap of the HP phase gradually reduces upon compression and reaches to a relatively small value of ~1.5 eV at 40 GPa. The band gap crossover at ~11 GPa is an indicative of the semiconductor-to- semiconductor transition. Moreover, the color changes in CrCl3 under compression further corroborate this electronic transition (Supplementary Fig. 7). Obviously, the isostructural transition is coupled with the semiconductor-to-semiconductor electronic transition. Interestingly, similar to the isostructural transition (Supplementary Figs. 2 and 4), the electronic transition in $CrCl_3$ is also reversible (Supplementary Figs. 5-6).

To reveal the underlying physics behind the experimentally observed isostructural phase transition, we have performed *ab initio* calculations on the monoclinic phase of $CrCl_3$. The calculated total energy as function of volume for three possible magnetic states (NM, FM, and AFM) of $CrCl_3$ is presented in Fig. 5a, and their lattice constants, relative total energies, and magnetic moments at the respective equilibrium volumes are listed in Table 1. At zero pressure, the FM state has the lowest relative total energy among the three candidates and it thus becomes the ground state. This is consistent with experimental observation that the intraplanar FM interactions dominate the magnetic behavior in the PM state[24]. Energetically, the NM phase is the most unfavorable with a energy of 1.548 (1.536) eV/formula higher than the FM (AFM) phase at their respective equilibrium volumes. Interestingly, there is a crossing between the two energetically competitive phases (i.e., FM and AFM), suggesting that monoclinic $CrCl_3$ can undergo a pressure-induced phase transition from the FM to the AFM ordering. To further support this, we have plotted the pressure versus enthalpy curves of the FM and AFM phases of $CrCl_3$. As shown in Fig. 5b, a phase transition takes place at ~5 GPa. Such a pressure-driven switching of magnetism from the FM to the AFM state should be a manifestation of the experimentally observed isostructural phase transition. We also notice that there is a difference between the predicted and experimental transition pressures. This is because our density functional theory (DFT) calculations do not account for the vdW interactions, but the overall trend of magnetic ordering crossover is unmistakable.

A compelling support for this argument can be obtained from the electronic structure calculations. The total and projected density of states (DOS) and band structures for the FM (at 0 GPa) and AFM (at 25 GPa) phases of monoclinic $CrCl_3$ are depicted in Fig. 6. For the FM phase, our calculation show that $CrCl_3$ is an indirect semiconductor with a band gap of 1.37 eV at zero pressure, which is slightly smaller than the value estimated from the above optical absorption measurements. This is reasonable since the DFT calculations typically underestimate the band gap of semiconductors and insulators[50]. When the pressure is increased, the band gap of the FM phase decreases gradually. Above 17.5 GPa, the spin-polarized calculations for the FM phase always converge to the NM solution (Fig. 5) and at the same time its band gap closes (the inset of Fig. 4b). However, as shown in the bottom panels in Fig. 6, the AFM phase still remains semiconducting with a band gap of 1.03 eV at the pressure of 25 GPa. Hence, the consideration of the FM state alone cannot explain the sudden rise in the experimental band-gap measurement. In fact, it is the magnetic crossover from the FM to the AFM ordering that hinders the band-gap closuring under pressure. Generally, an inverse correlation between pressure and bang gap has been expected in layered materials such as $MoS_2$[36] and $Sb_2Se_3$[51]. In contrast to $MoS_2$, for which an electronic transition from a semiconductor to a metal arises under pressure, $CrCl_3$ undergoes an anomalous semiconductor-to-semiconductor transition that is accompanied by a simultaneous magnetic transition from the FM to the AFM ordering. To the best of our knowledge, this is the first report that a layered TM compound is observed to demonstrate simultaneous isostructural, semiconductor-to-semiconductor, and magnetic phase transitions.

A localized versus collective electron model, first proposed by Goodenough[52], is very useful for understanding of the pressure-induced magnetic crossover in $CrCl_3$. Recently, it has been discovered that in 2D magnetic materials the interlayer exchange coupling strongly depends on the layer separation, and the layer stacking arrangement can alter the sign of the magnetic exchange and can thus modify the magnetic ordering.[53-56] As discussed above, the isostructural transition takes place due to interlayer rearrangements without the destruction of interlayer octahedral units. This isostructural transition causes the reduction of the interlayer spacing, which would result in the change of the spin polarizations, according to the Pauli's exclusion principle. Further theoretical explanation is needed to understand the detailed interaction between the FM and AFM exchanges with the narrowing of interlayer upon compression.

In summary, we have documented a pressure-induced isostructural transition in layered $CrCl_3$, leading to an unusual semiconductor-to-semiconductor transition. Concurrently, the magnetism of layered $CrCl_3$ is switched from the FM to AFM ordering. This study demonstrates that the pressure is a powerful control tool to tune the band gaps and magnetic properties of the layered TM compounds. Our findings would open up a new window for studying the tunable structural, magnetic, and electronic properties in other layered magnetic materials that are important for the applications in the electronic and spintronic devices.

**Methods**

**Sample characterization.** Single-crystal samples were purchased from Alfa Aesar with 99.9 % purity. At room temperature, the neutron diffraction experiment for $CrCl_3$ was carried out at the neutron beamline of the China Mianyang Research Reactor (CMRR), Minyang, China. The wavelength of the neutron beam was λ=1.59080 Å. The neutron diffraction data were analyzed using the Rietveld method[57] and FullProf program[58].

**High-pressure experiment measurements.** High-pressure measurements were performed using symmetric diamond-anvil cells with a cullet size of ~400 *μ*m in diameter (Supplementary Fig. 8). *In-situ* HP angle-dispersive XRD measurements were performed at the 4W2 beamline of the Beijing Synchrotron Radiation Facility (BSRF), Beijing, China. The incident X-ray wavelength was λ= 0.6199 Å. High-pressure Raman and PL measurements were performed with a Horiba LabRAM HR Evolution Raman system and the excitation wavelength of 532 nm from a DPSS laser was selected. To deduce mode frequencies and widths, Lorentzian fits were performed to each Raman spectra. High-pressure electrical resistance measurement for $CrCl_3$ was performed using the standard four-probe method. Four electrodes were cut from gold wire and cubic boron nitride (cBN) was employed as an insulating layer. In each HP experiment, single-crystal or powdered $CrCl_3$ sample was loaded into the hole drilled in a stainless-steel gasket pre–indented to ~30 mm thickness. A few ruby balls were also loaded in the sample chamber and severed as internal pressure calibrant.[59] Considering the fact that layered $CrCl_3$ is soft enough to offer quasi-hydrostatic conditions to itself in HP experiments, no pressure transmitting medium was used. The experiments were repeated using silicone oil as pressure-transmitting medium and similar results were obtained.

*Ab initio* **calculations.** The calculations were carried out within the framework of the spin-polarized DFT using the Perdew-Burke-Ernzerhof generalized gradient approximation[60] for the exchange-correlation functional as implemented in the Vienna *ab initio* simulation package (VASP)[61]. The all-electron projector augmented wave (PAW) method[62] and was adopted with $3s^23p^5$ and $3d^54s^1$ treated as valence electrons for Cl and Cr atoms, respectively. A plane-wave basis set with a large cutoff energy of 500 eV and dense *k*-meshes were employed for the considered phases to ensure that the numerical accuracy can resolve an energy difference of ~1 meV/atom. Forces on the ions were calculated through the Hellmann-Feynman theorem, allowing a full geometry optimization of different magnetic phases (i.e., NM, FM, and AFM) of $CrCl_3$.

**Acknowledgements**

The authors acknowledge financial support by the Shenzhen Peacock Plan (Grant No. KQTD2016053019134356), the Guangdong Innovative & Entrepreneurial Research Team Program (Grant No. 2016ZT06C279), the Key Research Platforms and Research Projects of Universities in Guangdong Province (Grant No. 2018KZDXM062), the Shenzhen Development and Reform Commission Foundation for Novel Nano-Material Sciences. We also acknowledge the support of the National Natural Science Foundation of China (Grants Nos. 51671126 and 11027405) and the Science Challenging Program (Grant No. JCKY2016212A501). This work was partially supported by the Shenzhen Development and Reform Commission Foundation for Shenzhen Engineering Research Center of Frontier Materials Synthesis at High Pressure. The synchrotron XRD measurements were performed at Beijing Synchrotron Radiation Facility (BSRF) in China. The Raman and PL spectroscopy measurements were performed at Materials Characterization and Preparation Center (MCPC), Southern University of Science and Technology (SUSTech). The help of Zhenwu Liao with performing Raman and PL experiments is gratefully acknowledged..

**Author contributions**

S.W., Y.L. and A.S.A. conceived the project. A.S.A performed experiments. Y.L. performed first-principles calculations. A.S.A., Y.L. and S.W. co-wrote the paper. All authors-A.S.A., Y.L., M.D., X.Z., L.F., Y.X., J.D., X.Y., X.Y., G.Z., Y.Z., S.W.- discussed the results.

**Additional Information**

Supplementary Information accompanies this paper.

**Competing interest:** The authors declare that they have no competing financial interests.

**Table and Figures:**

**Table 1.** Experimental and calculated lattice parameters [$a$ (Å), $b$ (Å), $c$ (Å), $\beta$ or $\gamma$ (°)], calculated relative total energy $E$ (eV/formula) and local magnetic moment $M$ ($\mu_B$/Cr) for three possible magnetic states (NM, FM and AFM) of $CrCl_3$ at zero pressure. The energy of the NM state is set as the reference energy (i.e., set to zero).

|       | $CrCl_3$ | $a$ | $b$ | $c$ | $\beta$ or $\gamma$ | $E$ | $M$ |
|-------|----------|--------|---------|---------|---------|--------|-------|
| Expt. | $R\text{-}3$ (100 K) | 5.9466 | 5.9466 | 17.2597 | 120 | - | - |
|       | $C2/m$ (293K) | 5.9600 | 10.3186 | 6.1164 | 108.54 | - | - |
| Calc. | $C2/m$ (NM) | 6.0056 | 10.3996 | 6.8341 | 108.45 | 0 | 0 |
|       | $C2/m$ (FM) | 6.0562 | 10.4886 | 6.9572 | 108.38 | -1.548 | 2.938 |
|       | $C2/m$ (AFM) | 6.0351 | 10.4683 | 6.8479 | 108.29 | -1.536 | 2.890 |

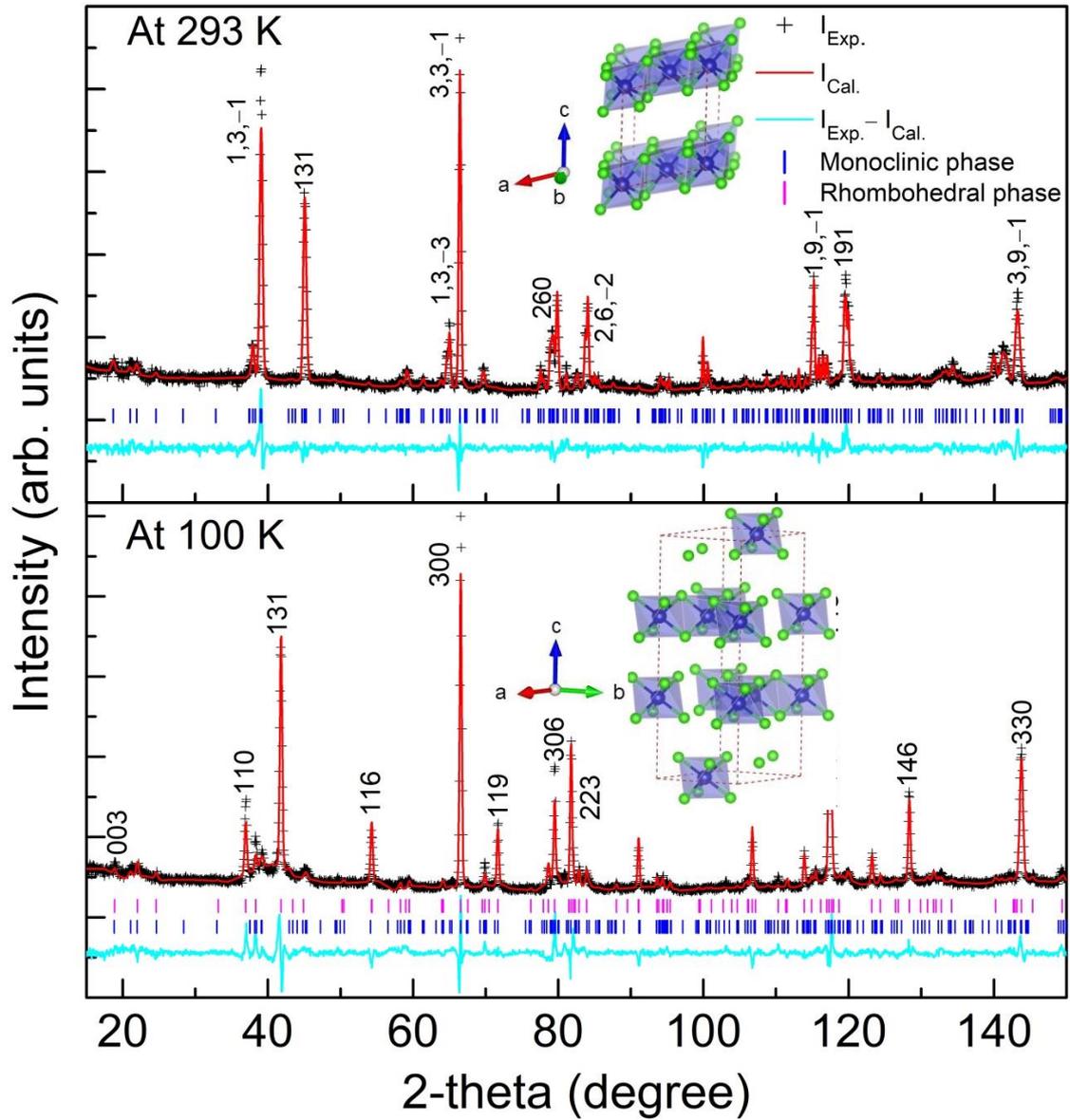

**Fig. 1** Neutron diffraction measurements for CrCl$_3$. Selected neutron diffraction pattern of CrCl$_3$ at 293 K (top panel) 100 K (bottom panel). Insets are the polyhedral views of the monoclinic and rhombohedral crystal structures, and the blue and green spheres represent Cr and Cl atoms, respectively.

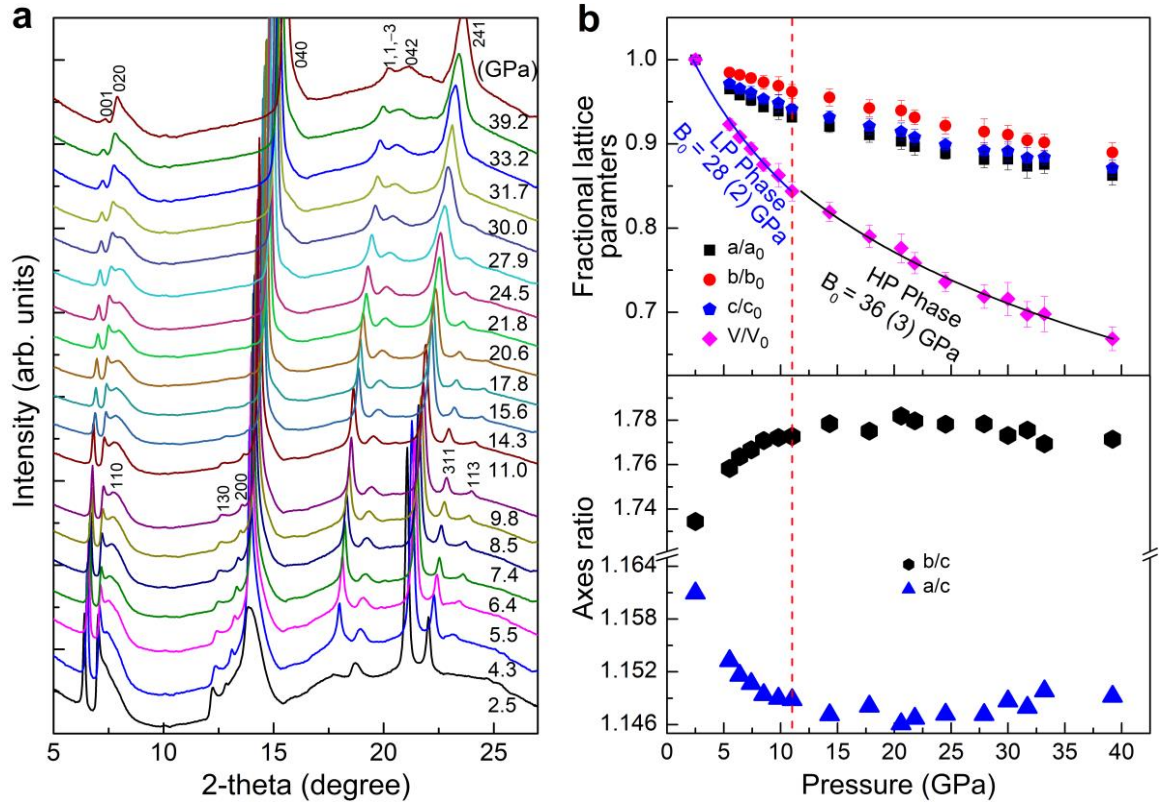

**Fig. 2** High-pressure synchrotron XRD measurements for $CrCl_3$. **a** Synchrotron XRD patterns of $CrCl_3$ at various selected pressures at 300K. **b** Fractional lattice parameters (i.e., $a/a_0$, $b/b_0$, $c/c_0$, and $V/V_0$) and axis ratios (i.e., $a/c$ and $b/c$) plotted as a function of pressure. Pressure-volume data are fitted to the second-order Birch-Murnaghan equation of states. The vertical dashed red line serves as a visual guide.

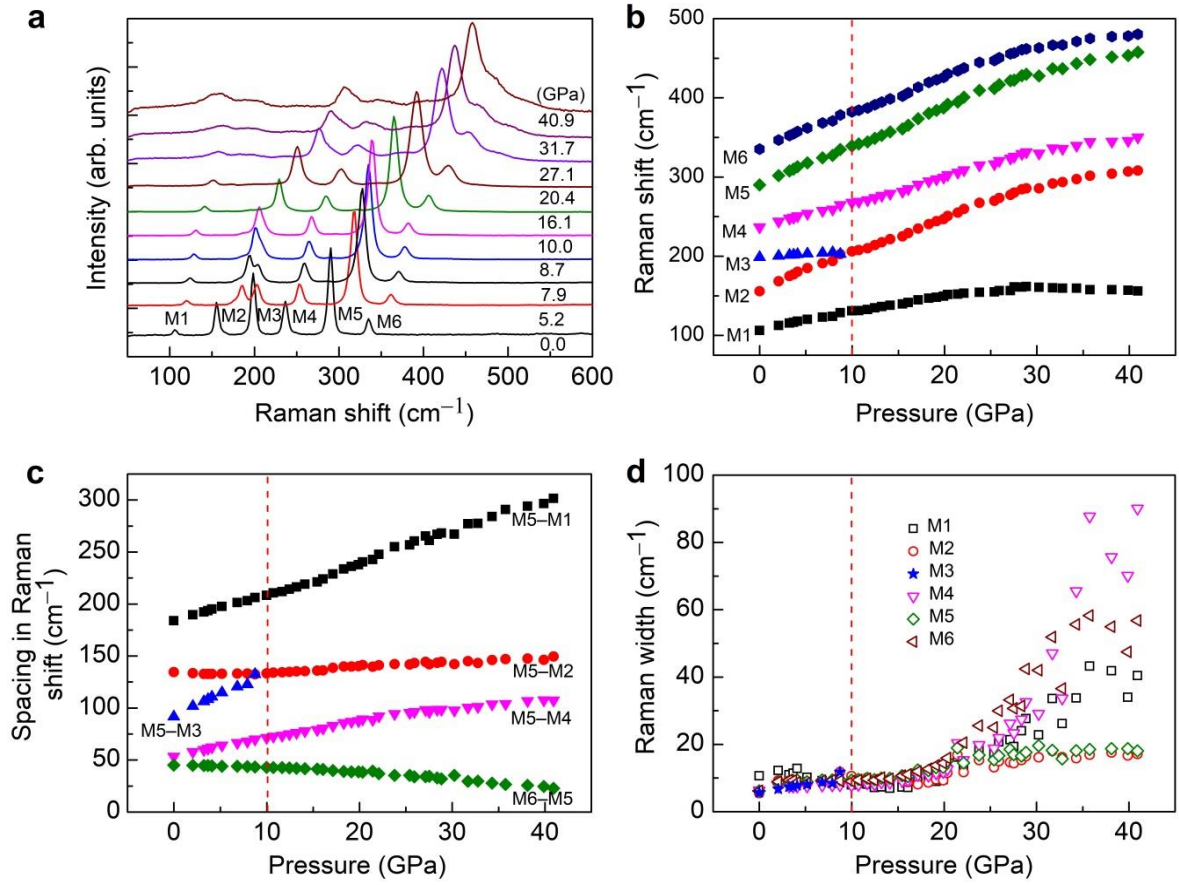

**Fig. 3** High-pressure Raman spectra of CrCl$_3$. **a** Raman spectra collected at selected pressures and room temperature. **b** Phonon frequency versus pressure. **c** Pressure-dependence of Raman peak separations. **d** Peak width as a function of pressure. The vertical dashed red lines serve as visual guides.

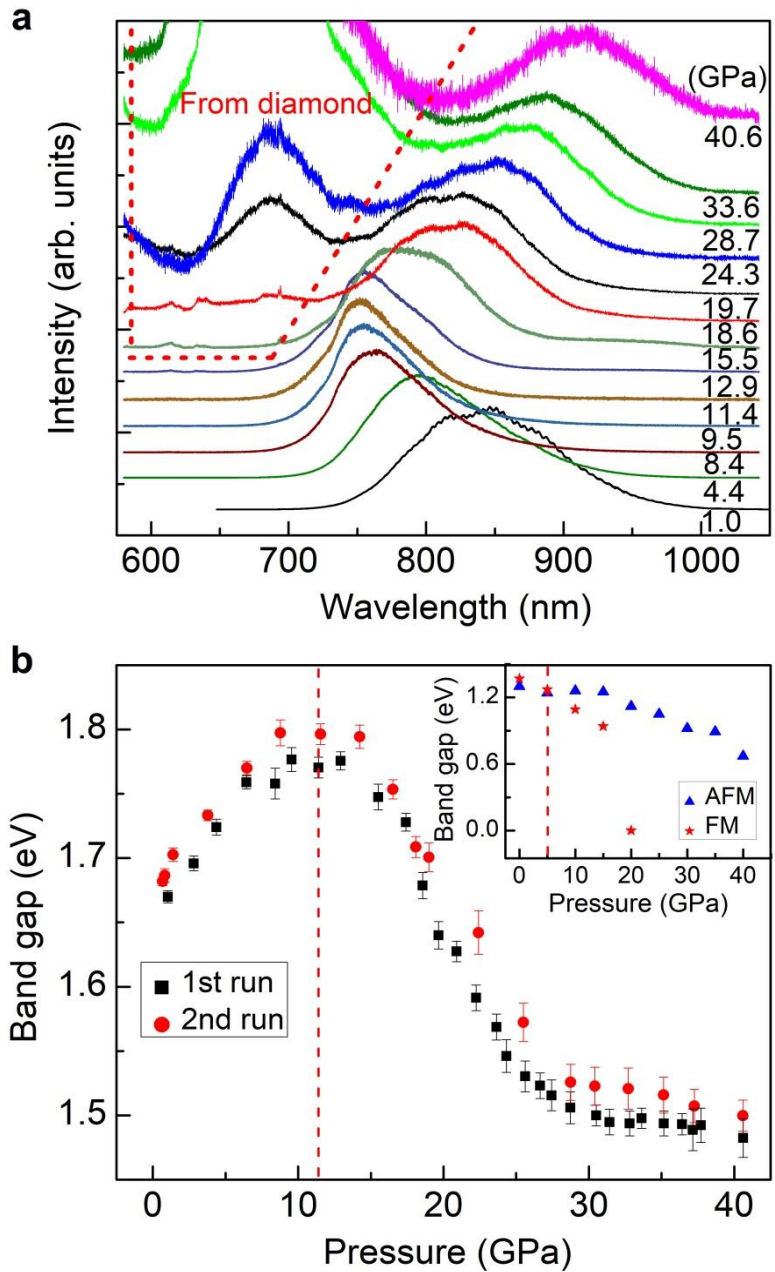

**Fig. 4** High-pressure PL spectroscopy measurements for CrCl$_3$. **a** Photoluminescence spectra collected at selected pressures and room temperature. Above ~15.5 GPa, background signals from diamond become significant and are marked within the dashed lines. **b** Derived optical band gap as a function of pressure. The closed squares and circles represent different runs of experiments. Inset: theoretically calculated pressure-dependent band gap for FM and AFM phases. The vertical dashed red line serves as a visual guide.

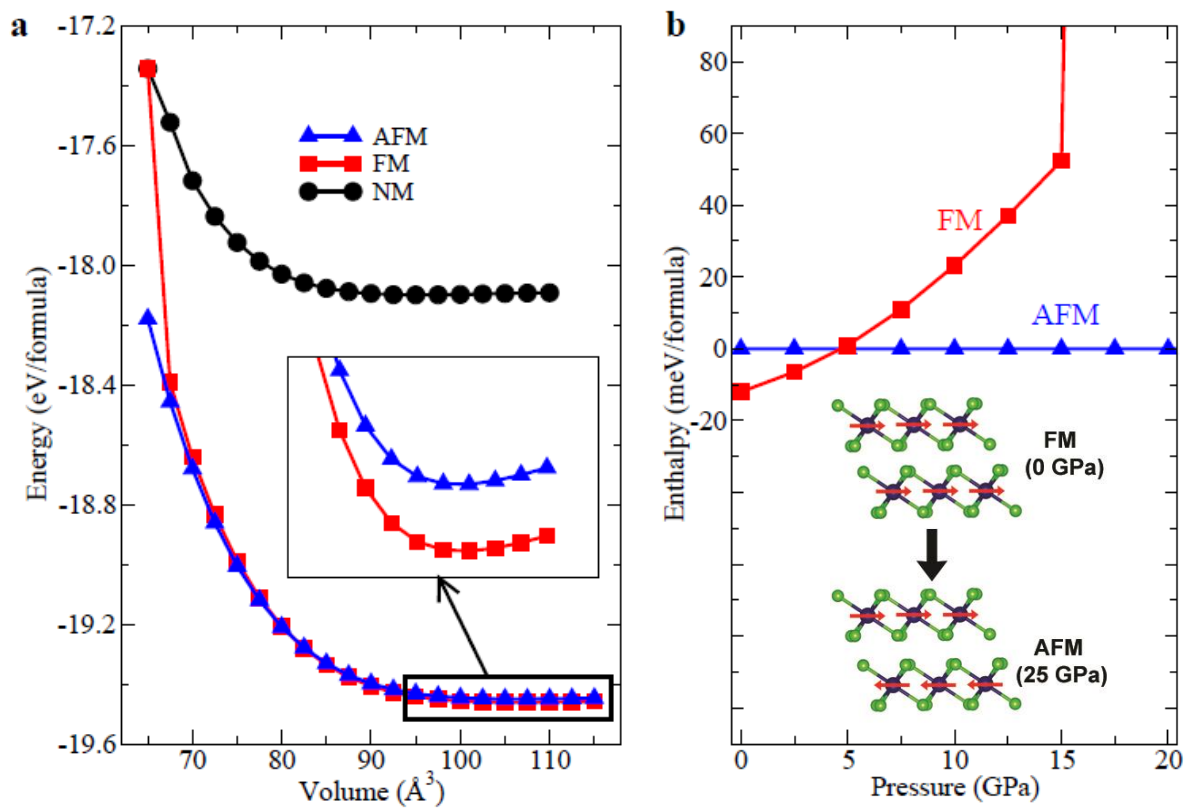

**Fig. 5** Theoretical calculations of the total energy and enthalpy of $CrCl_3$. **a** Calculated total energy versus volume of different magnetic phases (i.e., NM, FM, and AFM). **b** Calculated enthalpy versus pressure for the FM and AFM phases. The enthalpy of the AFM phase is set as the reference energy (i.e., set to zero) and all enthalpies are rescaled for one $CrCl_3$ formula. Schematic representation of magnetic transition from the FM to the AFM ordering is inset in b.

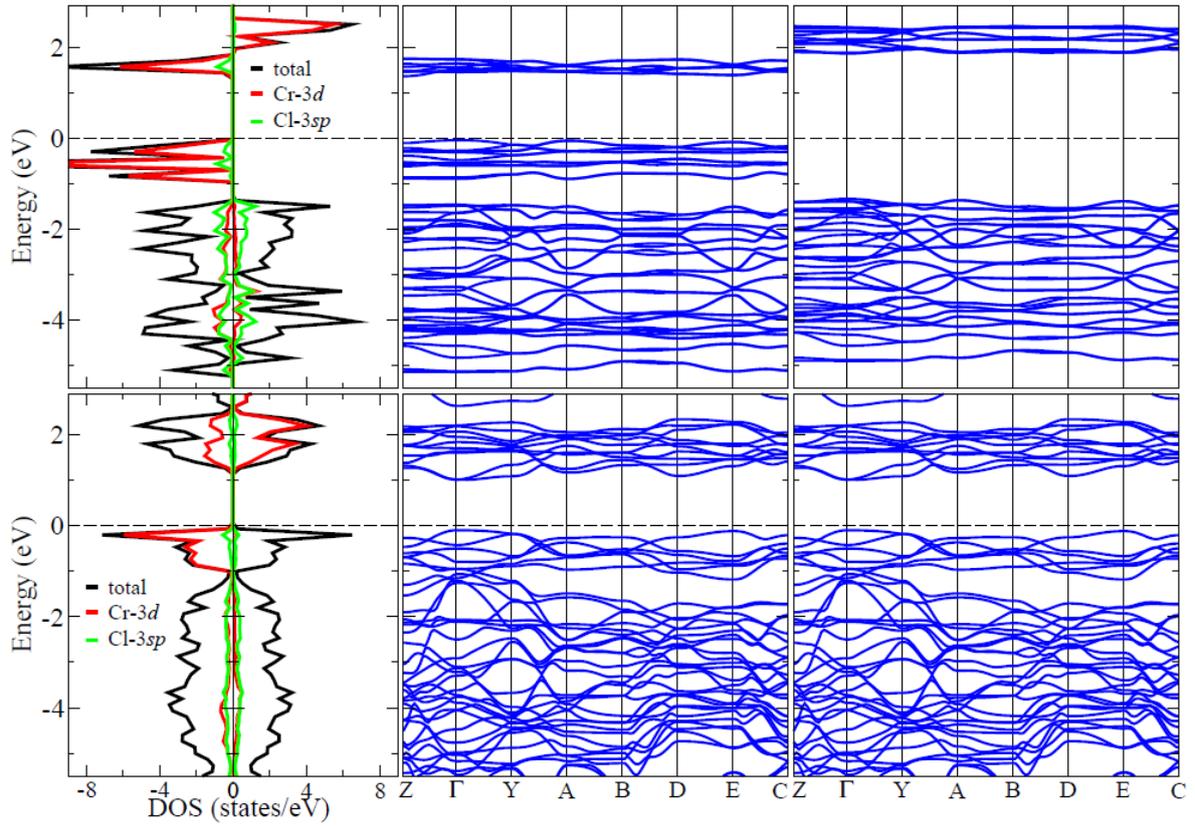

**Fig. 6** Theoretical calculations of the DOS and band structures of CrCl$_3$. Total and projected DOS and band structures of the FM phase at $P=0$ GPa (top panels) and the AFM phase at $P=25$ GPa (bottom panels). Their left and right panels represent the majority and minority spins, respectively. The Fermi levels are located at 0 eV as indicated by the horizontal dashed lines.